\documentclass[aps,prd,twocolumn,groupedaddress,showpacs,preprintnumbers,floatfix,nofootinbib]{revtex4}
\usepackage{graphicx}
\usepackage{amsmath}

\newcommand{\apjl}{Astrophys. J. Lett.}
\newcommand{\aap}{Astron. Astrophys.}
\newcommand{\aj}{Astron. J.}
\newcommand{\mnras}{Mon. Not. R. Astron. Soc.}
\newcommand{\pasp}{Publ. Astron. Soc. Pac.}
\newcommand{\lsim}{\mathrel{\hbox{\rlap{\lower.55ex\hbox{$\sim$}} \kern-.3em \raise.4ex \hbox{$<$}}}}
\newcommand{\gsim}{\mathrel{\hbox{\rlap{\lower.55ex\hbox{$\sim$}} \kern-.3em \raise.4ex \hbox{$>$}}}}

\begin{document}


\title{A Telescope Search for Decaying Relic Axions}


\author{Daniel Grin$^1$, Giovanni Covone$^{2}$, Jean-Paul Kneib$^{3}$, Marc Kamionkowski$^1$, Andrew Blain$^1$, and Eric Jullo$^4$}
\affiliation{$^1$California Institute of Technology, Mail Code 130-33, Pasadena, CA 91125}
\affiliation{$^2$INAF - Osservatorio Astronomico di Capodimonte, Naples, Italy}
\affiliation{$^3$Laboratoire d'Astrophysique de Marseilles, Traverse du Siphon, 13012 Marseilles, France}
\affiliation{$^4$European Southern Observatory, Santiago, Chile}

\begin{abstract}
A search for optical line emission from the two-photon decay of
relic axions was conducted in the galaxy clusters Abell $2667$
and $2390$, using spectra from the VIMOS (Visible Multi-Object
Spectrograph) integral field unit at the Very Large
Telescope. New upper limits to the two-photon coupling of the
axion are derived, and are at least a factor of $3$ more
stringent than previous upper limits in this mass window. The
improvement follows from larger collecting area, integration
time, and spatial resolution, as well as from improvements in
signal to noise and sky subtraction made possible by
strong-lensing mass models of these clusters. The new limits
either require that the two-photon coupling of the axion be
extremely weak or that the axion mass window between $4.5$ eV
and $7.7$ eV be closed. Implications for
sterile-neutrino dark matter are briefly discussed also.
\end{abstract}

\maketitle

\section{Introduction}
Axions are an obvious dark-matter candidate in some of the most conservative extensions of the standard model of particle physics. The magnitude of the charge-parity (CP) violating term
in QCD is tightly constrained by experimental limits to the electric dipole moment of the neutron, presenting the strong CP problem \cite{leader,baluni,crewther,harris}. Fine tuning can be avoided through the Peccei-Quinn (PQ) mechanism, in which a new symmetry (the Peccei-Quinn symmetry)
is introduced, along with a new pseudoscalar particle, the axion. These ingredients dynamically drive the CP violating term to zero \cite{pq,Raffelt,kt}. Via mixing with pions, axions pick up a mass, which is set by the PQ scale \cite{Raffelt}.

Below a mass of $10^{-2}~{\rm eV}$, axions will be produced through coherent oscillations of the PQ pseudoscalar, yielding a population of cold relics that dominate the dark-matter density \cite{notinvisible,kt,torig}. Above this mass, axions will be in thermal equilibrium at early times \cite{notinvisible,kt}. Unless $m_{{\rm a}}\gsim15~{\rm eV}$, the resulting relic density is insufficient to account for all the dark  matter, but high enough that axions will be a nontrivial fraction of the dark matter \cite{kt}. In either case, axions might be detectable through their couplings to standard-model particles.

The couplings of the axion are set by the PQ scale and the specific axion model \cite{kaplan,kt,Raffelt}. In the Dine-Fischler-Srednicki-Zhitnitski (DFSZ) axion model, standard-model fermions carry PQ charge, and so axions couple to photon pairs both via electrically charged standard-model leptons and via mixing with pions \cite{dine,zhitnitsky}. In hadronic axion models, such as the Kim-Shifman-Vainshtein-Zakharov (KSVZ) axion model \cite{kim79,svz80}, axions do not couple to standard-model quarks or leptons at tree level. In KSVZ models, axions couple to gluons through triangle diagrams involving exotic fermions, to pions via gluons, and to photons via mixing with pions. 

Constraints to the two-hadron couplings of axions come from stellar evolution arguments, from the duration of the neutrino burst from ${\rm SN}1987{\rm A}$, and from the upper limit to their cosmological density \cite{dicuskolb,raffelt2,raffelt3,sn1987a,kt,Raffelt,rglob}. Upper limits to the two-photon coupling of the axion come from searches for solar axions \cite{zioutas}, from upper limits to the intensity of the diffuse extragalactic background radiation (DEBRA) \cite{debra,debra2}, and from upper limits to x-ray and optical emission by galaxies and clusters of galaxies \cite{ressellt,notinvisible,Bershady:1990sw}. Recent searches for vacuum birefringence report evidence for the existence of a light boson \cite{zavattini1,zavattini2,zavattini3,anselm,gasperini,raffelt4,chris}, though in a region 
of parameter space already constrained by null solar axion searches \cite{zioutas,massoredondo,jainmandal}.

The two-photon coupling of the axion will lead to monochromatic line emission from axion decays to photon pairs. Although the lifetime of the axion is much longer than a Hubble time, the dark-matter density in a galaxy cluster is sufficiently high that optical line emission due to the decay of cluster axions
could be detected. This line emission should trace the density
profile of the galaxy cluster. Telescope searches for this
emission were first suggested in Ref. \cite{kephartw}. In
Ref. \cite{notinvisible}, this suggestion was extended to
thermally produced axions. A telescope search for this emission
was first attempted in Refs. \cite{ressellt,Bershady:1990sw}, in
which a null search imposed upper limits to the two-photon
coupling of the axion in the mass window $3~{\rm eV}\leq m_{{\rm
a}}\leq8~{\rm eV}$. Less stringent constraints have been obtained in searches for decaying galactic axions\cite{gnedin,ressellt}.

In the past few years, high-precision cosmic microwave background (CMB) and
large-scale-structure (LSS) measurements have become available and
allowed new constraints to axion parameters in this mass range.
In particular, axions in the few-eV mass range behave like hot
dark matter and suppress small-scale structure in a manner
much like neutrinos of comparable masses.  Reference ~\cite{hanraffelt}
shows that such arguments lead to an axion-mass bound  $m_{\rm
a}\leq 1.05$~eV.  Still, given uncertainties and
model dependences, it is important in cosmology to have several
techniques as verification.  For example, in extended, low-temperature ($\sim {\rm MeV}$) reheating models \cite{giudice1,giudice2,kawasaki2}, light relics like axions and neutrinos are produced non-thermally and have suppressed abundances,  evading CMB/LSS bounds, but may still show up in telescope searches for axion decay
lines \cite{grinprogress}.  Finally, other dark-matter
candidates may show up in such searches; the sterile
neutrino \cite{dodwid,dolgovhansen,shifuller,shaposh} 
is one example, which we will discuss below.  We are thus
motivated to re-visit the searches of
Refs. \cite{ressellt,Bershady:1990sw} and see whether new
telescopes, techniques, and observations may yield improvements.

Refs. \cite{ressellt,Bershady:1990sw} preceded the advent of
observations of gravitational lensing by galaxy clusters,
however, and so the cluster mass density profiles assumed were
not measured directly, but derived using x-ray data and
assumptions about the dynamical state of the clusters. The
constraints reported in Refs. \cite{ressellt,Bershady:1990sw}
depend on these assumptions. Today, gravitational-lensing data
can be used to determine cluster density profiles, independent
of dynamical assumptions \cite{kneib1996}. Thus, by using
lensing mass maps and by applying the larger collecting areas of
modern telescopes, cluster constraints to axions can both be
tightened, and made robust. The high spatial resolution of
integral field spectroscopy allows the use of lensing mass maps
to extract the component of intracluster emission that traces
the cluster mass profile. Cluster mass models can be used to
derive an optimal spatial weighting of the data, thus focusing
on parts of the cluster where the highest signal is expected.

To this end, we have conducted a search for optical line emission from the two-photon decays of thermally produced axions\footnote{Based on observations made with ESO
Telescopes at the Paranal Observatories (program ID 71.A-3010),
and on observations made with the NASA/ESA Hubble Space
Telescope, obtained from the data archive at the Space Telescope
Institute. STScI is operated by the association of Universities
for Research in Astronomy, Inc. under the NASA contract
NAS5-26555.}. We used spectra of the galaxy clusters Abell 2667 (A2667) and Abell 2390 (A2390) obtained with the Visible Multi-Object Spectrograph (VIMOS) Integral Field Unit (IFU), which has the largest field of view of any instrument in its class \cite{lefevre}. VIMOS is a spectrograph mounted at the third unit (M\'{e}lipal) of the Very Large Telescope (VLT), part of the European Southern Observatory (ESO) in Chile \cite{zanichelli}. In our analysis, we use mass models of the clusters derived from strong-lensing data, obtained with the Hubble Space Telescope (HST), using the Wide Field Planetary Camera \#2 (WFPC-2). 

We obtain new upper limits to the two-photon coupling of the axion in the mass window $4.5~{\rm eV}\leq m_{{\rm a}}\leq 7.7~{\rm eV}$ (set by the usable wavelength range of the VIMOS IFU) of $\xi\leq 0.003-0.017$. The two-photon coupling of the axion, $\xi$, is defined in Eq. (\ref{tau}) and discussed in Section \ref{thesec}. Although we search a smaller axion mass range than Refs. \cite{ressellt,Bershady:1990sw}, our upper limits improve on past work by a factor of $2.1-7.1$, depending on the candidate axion mass and how the limits of Ref. \cite{ressellt,Bershady:1990sw} are rescaled to correct for today's best-fit cosmological parameters and more accurate cluster mass profiles. Our data rule out the canonical KSVZ and DFSZ models in the $4.5~{\rm eV}-7.7~\rm{eV}$ window. However, \textit{theoretical} uncertainties in quark masses and pion couplings may allow for a much wider range of values of $\xi$ than the canonical KSVZ and DFSZ models allow, as emphasized by Ref.~\cite{murayama}, thus motivating the search for axions with smaller values of $\xi$. 

A quick estimate shows that our level of improvement is not unexpected: The collecting area of the VLT is a factor of $(8.1/2.1)^{2}$ greater than the $2.1\rm{m}$ telescope at Kitt Peak National Observatory (KPNO) used in Ref. \cite{Bershady:1990sw}. Our integration time is a factor of $10.8~\mbox{ksec}/3.6~\mbox{ksec}$ greater. The IFU allows us to cover $3.4$ times as much of the field of view as the spectrographs used at KPNO. Thus we estimate that our collecting area should be a factor of $\sim160$ higher than that of Ref. \cite{Bershady:1990sw}. If there is no signal, and if we are noise limited, we would expect a constraint to flux that is a factor of $\simeq 13$ more stringent than that of Ref. \cite{Bershady:1990sw}, and, since $I_{\lambda}\propto \xi^{2}$, upper limits to $\xi$ that are $\simeq 3.5$ times more stringent than those reported in Ref.  \cite{Bershady:1990sw}. 

We begin by reviewing the relevant theory and proceed to describe our observations. We then summarize our data analysis technique. The new limits to axion parameter space are then discussed along with other constraints. We conclude by pointing out the potential of conducting such work with higher redshift clusters. For consistency with the assumptions used to derive the strong-lensing maps used in our analysis, we assume a $\Lambda$CDM cosmology parameterized by $h=0.71$, $\Omega_{{\rm m}}=0.30$, and $\Omega_{\Lambda}=0.70$, except where explicitly noted otherwise.
\section{Theory}
\label{thesec}
To predict the expected intensity of the optical signal due to axion decay, given the mass distribution of a galaxy cluster, we need to know the total mass density in axions. If $m_{\rm a}\geq 10^{-2}~{\rm eV}$, standard thermal freeze-out arguments show that the mass density of thermal axions today is\begin{equation}\Omega_{{\rm a}}h^{2}=\frac{m_{{\rm a},{\rm eV}}}{130}\label{thermalfreeze},\end{equation} where $m_{{\rm a},{\rm eV}}$ is the axion mass in units of ${\rm eV}$ \cite{kt,notinvisible,ressellt}. 

Thermal axions in our mass range of interest, which become nonrelativistic when $4.5~{\rm eV}\leq T\leq 7.7~{\rm eV}$, will have a velocity dispersion today of 
$
\langle v_{{\rm a}}^{2}/c^2\rangle^{1/2}= 4.9\times 10^{-4} m_{{\rm a},{\rm eV}}^{-1},$ low enough that axions will bind to galaxy clusters  \cite{notinvisible,ressellt}. The maximum axionic mass fraction $x_{{\rm a}}^{{\rm max}}$ of a cluster is \cite{gt,ressellt}
\begin{equation}
x_{{\rm a}}^{{\rm max}}=1.2\times 10^{-2} m_{{\rm a},{\rm eV}}^{4} a_{250}^{2}\sigma_{1000},\label{xamaxa}
\end{equation}
where $\sigma_{1000}$ is the velocity dispersion of the cluster in units of $1000~ {\rm km}~{\rm s}^{-1}$, and $a_{250}$ is the cluster core radius in units of $250~h^{-1}~{\rm kpc}$. For $m_{{\rm a},{\rm eV}}>3$ and typical cluster values, there is ample phase space to accommodate an axionic mass fraction of $x_{{\rm a}}=\Omega_{{\rm a}}/\Omega_{{\rm m}}$.

Axions decay to two photons via two channels. In one, axions couple to neutral pions through their two-gluon coupling and 
two QCD vertices. These pions then decay to photon pairs through QED triangle anomaly diagrams. In the other mechanism, axions couple directly to standard-model fermions through triangle anomaly diagrams, which then couple to photon pairs through QED vertices. The total decay rate is derived by taking the sum of the matrix elements for these two processes, and then incorporating the relevant kinematic factors, yielding an axion lifetime of \cite{Raffelt,ressellt}
\begin{eqnarray}
\nonumber
\tau=6.8\times 10^{24} \xi^{-2} m_{{\rm a},{\rm eV}}^{-5}~{\rm s},\\
\mbox{where } \xi\equiv\frac{4}{3}\left(E/N-1.92\pm 0.08\right).\label{tau}\end{eqnarray}
The values of $E$ and $N$ depend on the axion model chosen, but by parameterizing $\tau$ in terms of $\xi$, we are able to obtain model independent upper limits to $\xi$. The negative sign comes from interference between the different channels for the two-photon decay of axions. The uncertainty in the theoretical value of $\xi$ comes from uncertainties in the quark masses and pion-decay constant, and may in fact be larger than indicated by Eq.~(\ref{tau}). A complete cancellation of the axion's two-photon coupling is possible for KSVZ models, in which $E/N=2$, and even for DFSZ axion models, in which $E/N=8/3$ \cite{murayama}. It is thus hasty to claim that an upper limit on $\xi$ truly rules out axions; it always pays to keep looking.

The rest-frame wavelength of the axion-decay line is
$\lambda_{{\rm a}}=24,800{{\rm \AA}}/m_{{\rm a},{\rm eV}}$, and
the line width is dominated by Doppler broadening.  If the axion
has a cosmological density given by Eq.~(\ref{thermalfreeze})
and its mass fraction in the cluster is $x_a$, then the
observer-frame specific intensity from axion decay is
\begin{widetext}
\begin{equation}
I_{\lambda_{o}}=2.68\times 10^{-18}\times \frac{m_{{\rm a},{\rm eV}}^{7} \xi^{2}\Sigma_{12}\exp{\left[-\left(\lambda_{r}-\lambda_{{\rm a}}\right)^{2}c^{2}/\left(2\lambda_{{\rm a}}^{2}\sigma^{2}\right)\right]}}{\sigma_{1000} (1+z_{{\rm cl}})^{4}S^{2}(z_{{\rm cl}})} {\rm cgs},
\label{lintens}
\end{equation}\end{widetext} 
where $\lambda_{{\rm o}}$ denotes wavelength in the observer's
rest frame, $\lambda_{r}=\lambda_{o}/(1+z_{{\rm cl}})$, cgs
denotes units of specific intensity (${\rm ergs}~{\rm
cm}^{-2}~{\rm s}^{-1}~{{\rm \AA}}^{-1}~\mbox{arcsec}^{-2}$),
$S(z_{{\rm cl}})\equiv d_{a}(z_{{\rm
cl}})/\left[c/\left(100~{\rm km}~{\rm s}^{-1}~{\rm
Mpc}^{-1}\right)\right]$ is a dimensionless angular-diameter
distance, and $\Sigma_{12} \equiv \Sigma/\left(10^{12} M_{\odot}
{\rm pixel}^{-2}\right)$ is the normalized surface mass density
of the cluster.  If for some reason (e.g., low-temperature reheating \cite{grinprogress}), the cosmological axion mass
density is lower than indicated by Eq.~(\ref{thermalfreeze}),
then the intensity in Eq.~(\ref{lintens}) is decreased accordingly.

The cluster mass density was determined by
fitting parameterized potentials to the locations of
gravitationally lensed arcs. In our mass maps, one pixel is
$\sim 0.5~\rm{arcsec}$ across. The intensity predicted by
Eq.~(\ref{lintens}) is comparable with that of the night-sky
continuum, and so it is crucial to obtain a good sky subtraction
when searching for an axion-decay line in clusters. Fortunately,
the spatial dependence of the cluster density and the expected
signal provides a natural way to separate the background from an
axion signal, as discussed in Section \ref{secanal}. 

\section{Observations}
\subsection{Imaging Data}
\begin{figure}
\includegraphics[width=80mm]{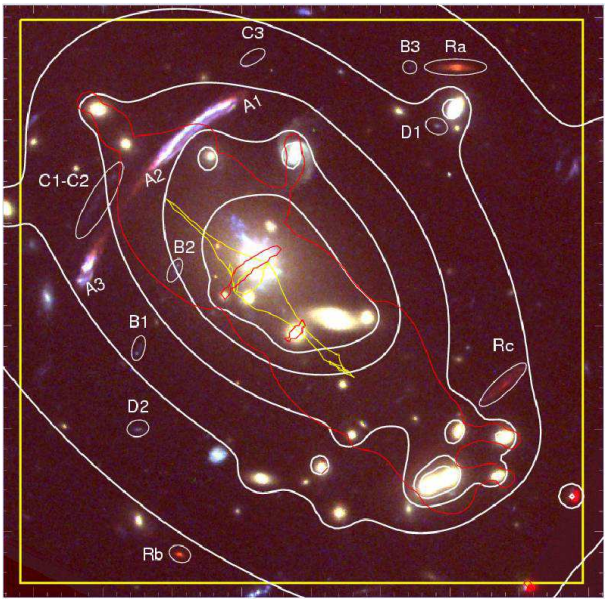}
\caption{Image of the Abell 2667 cluster core imaged with HST in the
F450W, F606W, and F814W filters. The white (thin yellow) square shows the IFU field of view, which is $54"\times 54"$. North is to the top and east is to the
left. Note the strongly magnified gravitational arc north-east of the central galaxy. 
The white curves correspond to
iso-mass contours from the lens model; the dark gray (red) line is the
critical line at the redshift of the giant arc. 
The field of view is centered on $\alpha_{J2000}$=23:52:28.4,
$\delta_{J2000}$=$-$26:05:08. At a redshift of $z=0.233$, the angular scale is $3.661$ kpc/arcsec.}
\label{lensfig}
\end{figure}

\begin{figure}
\includegraphics[width=80mm]{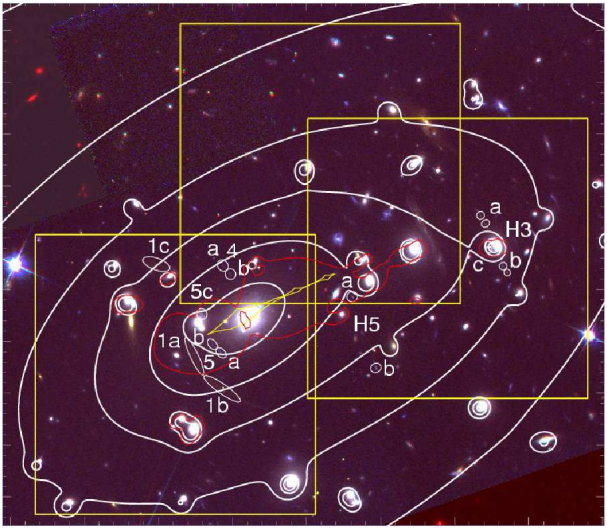}
\caption{Image of the Abell 2390 cluster core imaged with HST in the
F450W, F606W, and F814W filters. The white (thin yellow) squares correspond to the IFU field of view in different pointings. The white curves correspond to
iso-mass contours from the lens model. The dark gray (red) line is the critical line at the redshift of the giant arc, labelled 1. Each square is $54"\times 54"$. North is to the top and east is to the left. The field of view is centered on $\alpha_{J2000}$=21:53:36.970, $\delta_{J2000}$=+17:41:44.66. At a redshift of $z=0.228$, the angular scale is $3.601$ kpc/arcsec.}
\label{lensfig2}
\end{figure}

To construct the lensing models used in our analysis and to mask out IFU fibers corresponding to cluster galaxies and other bright sources, we used images of A2667 and A2390 obtained with the HST and the VLT.  The cluster A2667 was observed with HST on October 10-11, 2001, using WFPC-2 in the F450W, F606W, and F814W filters, with total exposure times of $12.0~{\rm ksec}$, $4.00~{\rm ksec}$, and $4.00~{\rm ksec}$, respectively \cite{covone}. The cluster A2390 was observed with HST on December 10, 1994, using WFPC-2 in the F555W and F814W filters and total exposure times of $8.40~{\rm ksec}$ and $10.5~{\rm ksec}$ \cite{jullo}. After pipeline processing, standard reduction routines were used with both clusters to combine the frames and remove cosmic rays. Figs.~\ref{lensfig} and \ref{lensfig2} are images of the cluster cores, with iso-mass contours overlaid from our best-fit lensing models. 


On May $30$ and June $1$, $2001$, near-infrared J-band and H-band observations of A2667 were obtained with ISAAC on the VLT \cite{covone}. The total exposure times for the J and H band ISAAC data were $7.93~{\rm ksec}$ and $6.53~{\rm ksec}$, respectively. The final seeing was $0.51{\rm ''}$ and $0.58{\rm ''}$ in the J and H bands, respectively.

\subsection{VIMOS Spectra}
The massive galaxy clusters A2667 and A2390 were observed with VIMOS, between June $27$ and $30$, 2003 \cite{covone,jullo}. The IFU is one of three modes available on VIMOS, and consists of 4 quadrants, each containing 1600 fibers. We used an instrumental setup in which each fiber covered a region of $0.67''$ in diameter. A single pointing covered a $54''\times 54''$ region of the sky. Roughly $10\%$ of the IFU field of view is unresponsive because of incomplete fiber coverage. A low resolution blue (LR-Blue) grism was used, covering the wavelength range $3500 {{\rm \AA}}$ to $7000 {{\rm \AA}}$ with spectral resolution $R\approx 250$ and dispersion $5.355{{\rm \AA}}/{\rm pixel}$. The FWHM of the axion-decay line is $195{\rm \AA}~ \sigma_{1000}~m_{{\rm a},{\rm eV}}^{-1}$, and so the LR-Blue grism can resolve this line, without spreading a faint signal over too many wavelength pixels. Unfortunately, because spectra from contiguous pseudo-slits (sets of 400 spectra) on the CCD overlap, the first and last $50$ pixels on most of the raw spectra are unusable, reducing the spectral range to $4000{\rm \AA}-6800{\rm \AA}$, corresponding to an axion mass-range of $4.5\leq m_{{\rm a},{\rm eV}}\leq 7.7$ at the nearly identical redshifts ($z \approx 0.23$) of the two clusters.  Further observational details are discussed in Ref. \cite{covone}.

The total exposure time for each cluster was $10.8~{\rm ksec}$ ($4\times 2.70~{\rm ksec}$ exposures). Calibration frames were obtained after each of the exposures, and a spectrophotometric standard star was observed. In order to compensate for the presence of a small set of bad fibers, we used an offset between consecutive exposures. At a redshift of $z=0.233$ (A2667), the IFU covers a physical region of $198~{\rm kpc}\times198~{\rm kpc}$ in the plane of the cluster. At a redshift of $z= 0.228$ (A2390), the IFU covers a physical region of $195~{\rm kpc}\times195~{\rm kpc}$. 

\subsection{Reduction of IFU data}
\label{initred}
If axions exist and are present in the halos of massive galaxy clusters, a distinct spectral feature will appear in VIMOS-IFU data. At a rest-frame wavelength $\lambda_{{\rm a}}$, we will observe a spatially extended emission line whose intensity traces the  projected dark-matter density. Revealing such a faint, spatially extended signal requires great care in correcting for fiber efficiency and in subtracting the sky background, because the instrument itself can impose spatial variation in the sky background through varying IFU fiber efficiency. 

The VIMOS-IFU data were reduced using the VIMOS Interactive Pipeline
Graphical Interface (VIPGI), and the authors' own routines \cite{vipgiscodeggio}. References \cite{zanichelli,covone} give both a detailed description of the methods and an assessment of the quality of VIPGI data reduction. The reduction steps that precede the final combination of the dithered exposures into a single data cube are performed on a quadrant by quadrant basis. The main steps are the following \cite{covone,dodorico,vipgiscodeggio,zanichelli,horne}: extract spectra from the raw CCD data at each pointing, calibrate wavelength, remove cosmic rays, determine fiber efficiencies, subtract the sky background, and calibrate flux. 

The exposures were bias subtracted. Cosmic-ray hits were removed with an efficient automatic routine based on a $\sigma$-clipping algorithm, which exploits the fact that cosmic-ray hits show strong spatial gradients on the CCD \cite{zanichelli}, in contrast to the smoother spatial behavior of genuine emission lines. In Ref. \cite{Bershady:1990sw}, spectra were obtained using a limited number of long-slit exposures, so the removal of a small number of incorrectly identified cosmic-ray hits could thwart a search for line emission from decaying axions. An axion-decay line, however, must smoothly track the density profile of the cluster. Our spectra are highly spatially-resolved, and so cosmic-ray hits can be removed safely using our cleaning algorithm. Using the raw CCD spectral traces, we verified that the signals removed by the cleaning algorithm bore the distinctive visual signatures of cosmic-ray hits.

VIPGI usually determines fiber efficiencies by normalizing to the flux of bright sky lines; this technique yielded data cubes with prominent bright and dark patches (each covering $\sim 20\times 20$ fibers). It is conceivable that an accidental correlation of these patches with the cluster density profile could lead to a spurious axion signal. To avoid this possibility, we measured fiber efficiency using high signal to noise continuum arc-lamp exposures (analogous to flat-fielding for images and henceforth referred to as flat-fielding). The resulting flat-fielded data cubes were much less patchy, and were thus used for all subsequent analysis.

The VIMOS IFU does not have a dedicated set of fibers to determine the sky-background level. VIPGI usually determines the sky statistically at each wavelength. VIPGI first groups the fibers in each quadrant into three sets according to the shape of a user selected sky-emission line, and then takes the statistical mode of the counts in each set and subtracts it from the counts measured in each fiber in the set \cite{vipgiscodeggio}. Although axions (and thus their decay luminosity) trace the centrally peaked density profile of the cluster, the average decay luminosity would be wiped out by this procedure, and lead to a spurious depression in measurements of $\xi$. The sky subtraction implemented in VIPGI is thus unsuitable for our axion search and was not applied. A customized sky-subtraction was applied, as discussed in Section \ref{extraction}.

Flux is calibrated separately for each IFU quadrant, using observations of a standard star. Finally, the four fully reduced exposures are combined. The final data cube for A2667 is made of 6806 spatial elements, each one containing a spectrum from $3500{\rm \AA}$ to $7000 {\rm \AA}$, and covers a sky area of 0.83 arcmin$^2$, centered $5$ arcsec south-west of the brightest cluster galaxy. The final data cube for A2390 is made of $24,645$ spatial elements, each one containing a spectrum from $3500{\rm \AA}$ to $7000 {\rm \AA} $, and covers a sky area of  $3.11$ arcmin$^2$, centered $15$ arcsec north-east of the brightest cluster galaxy. The median spectral resolution is $\simeq 18 \rm{\AA}$. For further discussion of the process used to generate the data cubes, see Refs. \cite{zanichelli,covone}.

After producing data cubes in VIPGI, we passed these data cubes to a secondary routine that searches for emission from axion decay and estimates the noise in our spectra. The most obvious source of error is Poisson counting noise. The number of photons observed at wavelength $\lambda$ in the $\rm{j}^{\rm{th}}$ spatial bin is just $N_{\lambda,j}=E\left[F_{\lambda,j}/(hc/\lambda) \right]\delta t \delta \lambda \delta A$, where $\delta A=51.2\rm{m}^{2}$ is the collecting area of the M\'{e}lipal telescope, $\delta \lambda=5.355\rm{\AA}$ is the dispersion of a single VIMOS spectral pixel, $\delta t$ is the integration time, $F_{\lambda,j}$ is the flux in the j$^{th}$ pixel at wavelength $\lambda$, and $E$ is the end to end mean efficiency of VIMOS mounted at M\'{e}lipal. The Poisson counting noise is $\delta N_{\lambda,j} \approx \sqrt{N_{\lambda,j}}$, and so $\delta I_{\lambda,j} \approx I_{\lambda,j}/\sqrt{N_{\lambda,j}}$. A secondary error source is flux contamination from neighboring pixels. To include this error, we use the $5\%$ estimate of Ref. \cite{zanichelli}, calculate the `leakage' contribution to noise at each pixel by taking the mean flux of all the nearest neighboring pixels, and multiply it by $5\%$. We also use time-logged measurements of the CCD bias and dark-current, with the appropriate integration time, to calculate the additional noise from these sources\footnote{\tt{http://www.eso.org/observing/dfo/quality/VIMOS/toc.html}\rm.}. Finally we estimate the flat-fielding noise using the rms difference between different sets of efficiency tables. These errors are added in quadrature to obtain a data cube of the estimated errors in specific intensity. 

In Refs. \cite{ressellt,Bershady:1990sw}, slit locations were chosen to avoid the locations of known galaxies, as well as regions that showed statistical evidence for faint galaxies \cite{pc}. Likewise, we masked out IFU fibers that fell on the locations of individual bright sources. Bright sources were identified in each cluster image by tagging pixels where the image intensity exceeded the median by more than $1\sigma$ and masking IFU fibers that fell on these pixels. Practically, this means that $40\%$ of the fibers in each data cube are left unmasked. The images used to generate this mask are broadband, and so this masking technique will not mask out an axion-decay signal. The accidental inclusion of galaxies could conceivably lead us to erroneously attribute their emission to axion decay. This is unlikely, given that the spectra of cluster galaxies are dominated by continuum emission and line absorption. If we see an indication of emission due to line decay, however, we may have to revise our masking criteria to take account of this possibility. As we shall see later, we imposed upper limits to axion decay, and can safely use the chosen masking criterion. The resulting masks were visually inspected to verify that most of the masked fibers fall near galaxies. To extract the density dependent component of the cluster spectra, we apply a mass map obtained from gravitational lensing observations.

\section{Analysis}
\subsection{Strong Lensing and Cluster Mass Maps}
To model the mass distribution of A2667 and A2390, we used both a 
cluster mass-scale component (representing the contribution of the dark-matter halo and the intracluster
medium) and cluster-galaxy mass components as in Refs. \cite{kneib1996,smith2005}. 
Cluster galaxies were selected according to their 
redshift (when available, in the inner cluster region
covered with VIMOS spectroscopy) or 
their color, thus selecting galaxies belonging to 
the cluster red sequence. For A2667, ISAAC images were used to determine J-H colors, whereas for A2390, HST images were used to determine I-V colors. The lensing contribution from more prominent foreground galaxies was also included, rescaling their lensing properties using the appropriate redshift.

All model components were parameterized using a smoothly truncated pseudo-isothermal mass distribution model (PIEMD) \cite{kassiolakovner1993}, which avoids both the unphysical
central-density singularity and the infinite spatial extent of the singular isothermal model.

The galaxy mass components were chosen to have the same position, ellipticity and orientation as their corresponding images.  The K-band luminosity of the galaxies was computed, assuming a typical E/S0 spectral energy distribution (redshifted but uncorrected for evolution of constituent stellar populations). Their masses were estimated using the K-band luminosity, calculated assuming a global mass to light
ratio ($M/L$) and the Faber-Jackson relation \cite{faberjackson}. The final mass model is made of 70 components, 
including the large scale cluster halo and the individual galaxies.

\begin{figure}
\includegraphics[width=80mm]{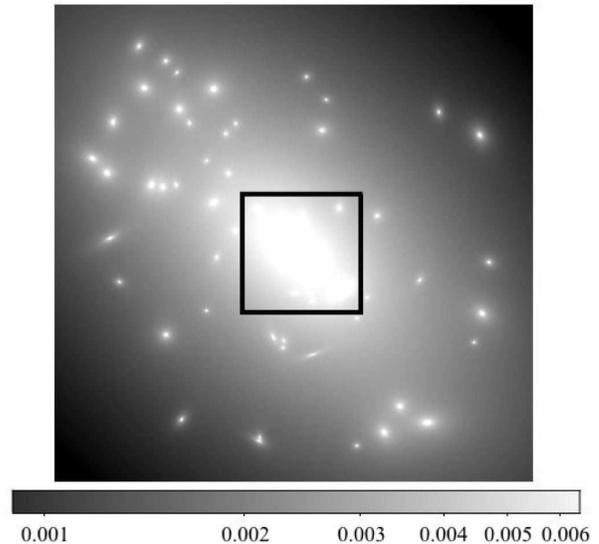}
\caption{Mass map of A2667. The intensity of the image scales with density (in units of $10^{12} M_{\odot}~\rm{pix}^{-2}$), where $1~\rm{pix}=0.50"$. A density scale is provided on the bottom of the image. The horizontal extent of this map is $222.6''$. The vertical extent is $200.0''$. The thick black line indicates the spatial extent of the IFU head on the mass map.}
\label{massmap2667}
\end{figure}

\begin{figure}
\includegraphics[width=80mm]{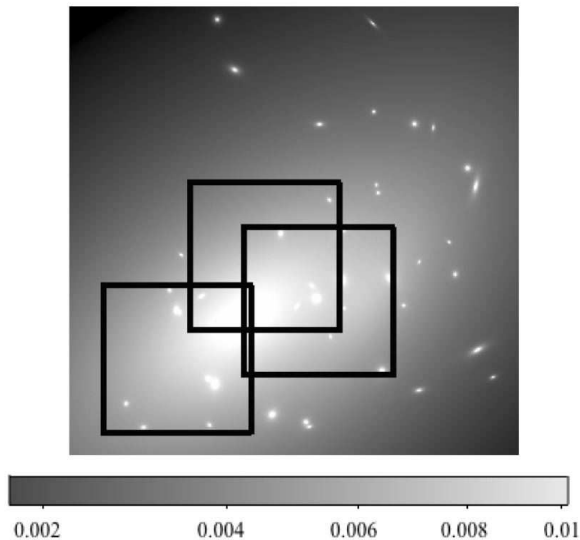}
\caption{Mass map of A2390. The intensity of the image scales with density (in units of $10^{12} M_{\odot}~\rm{pix}^{-2}$), where $1~\rm{pix}=0.50"$. A density scale is provided on the bottom of the image. The horizontal spatial extent of the map is $157.5"$. The vertical extent is $150.0''$. The thick black lines indicate IFU pointings used to construct our data cubes.}
\label{massmap2390}
\end{figure}

Using the {\tt LENSTOOL} ray-tracing code \cite{kneib1993} with the HST images, we iteratively implemented the constraints from the gravitational lenses. Lensing mass models with $\chi^2\leq 1$ were found by fitting the 
ellipticity, orientation, center, and mass parameters 
(velocity dispersion, core radius, and truncation radius) 
of the cluster scale component, as well as
the truncation radius and velocity dispersion
of the ensemble of cluster galaxies, using scaling relations for early-type galaxies \cite{natarajankneib1997}. Cluster galaxy redshifts were measured using the IFU data \cite{covone}. The bright central galaxy and several galaxies near the locations of strong-lensing arcs were modeled separately from the ensemble. The resulting cluster density maps for A2667 and A2390 are shown in Figs.~\ref{massmap2667} and \ref{massmap2390} \cite{covone,jullo}. Statistical errors in the mass model parameters were propagated through the relevant code to produce a fiber by fiber map of statistical errors in $\Sigma$. These maps were then used to weight different IFU fibers and thus maximize the signal to noise ratio of any putative line emission from axion decay.

\subsection{Extraction of One Dimensional Spectra}
\label{extraction}
\label{secanal}

Using density maps of A2667 and A2390, we can optimally weight averages over fibers to maximize the contribution from high density regions of the cluster. This maximizes the signal to noise ratio of our axion search by emphasizing IFU fibers where maximum signal from axion decay is expected. These maps allow us to separate emission correlated with the mass profile of the cluster, which could be due to axion decay, from a sky background that we assume to be spatially homogeneous. Our technique is an IFU generalization of the long-slit `on-off' sky-subtraction technique presented in Refs. \cite{ressellt,Bershady:1990sw}. The real sky background is certainly not perfectly homogeneous, but by making this assumption, we are being maximally conservative. With our reduction method, any density correlated spatial dependence in the sky background will show up as putative emission from axion decay. If evidence for a signal is seen, we will have to be careful to avoid being confused by sky line emission. The projected surface density of the cluster at the location in the lens plane associated with the $i^{\rm{th}}$ fiber is denoted $\Sigma_{12,i}$. Assuming that the only spatially dependent signal comes from axions, we can then model the actual intensity $I_{\lambda,i}^{\rm{mod}}$ at a given wavelength $\lambda$ and spatial pixel $i$ as $I_{\lambda,i}^{\rm{mod}}=\langle I_{\lambda}/\Sigma_{12}\rangle \Sigma_{12,i}+b_{\lambda}$, where $b_{\lambda}$ represents the contribution of a spatially homogeneous sky signal, and $I_{\lambda,i}$ is the specific intensity in the $i^{\rm{th}}$ fiber at wavelength $\lambda$. Since the signal from axion decay is bounded from above by the total component of the signal proportional to $\Sigma_{12,i}$, a measurement of $\langle I_{\lambda}/\Sigma_{12}\rangle$ will either provide evidence of axion decay, or impose an upper limit on $\langle I_{\lambda}/\Sigma_{12}\rangle_{\rm{axion}}$.  Using a simple linear fit to separate the sky background from signal, we extract the array $\langle I_{\lambda} /\Sigma_{12} \rangle$ from each cluster data cube. 

At a small number of wavelengths, this yielded negative (unphysical) values for one or both fitted parameters. To avoid this, we fit for $\langle I_{\lambda}/\Sigma_{12}\rangle$ and $b_{\lambda}$, subject to the obvious constraints $\langle I_{\lambda}/\Sigma_{12}\rangle\geq 0$ and $b_{\lambda}\geq 0$, estimating errors $\sigma_{\lambda,i}$ as described in Section \ref{initred}. Estimated errors in $\Sigma_{12}$ were also included in the fit. Residuals from the best fit are due to flux noise, imperfect masking of galaxies, and variations in fiber efficiency unaccounted for by the flat-fielding procedure. As can be seen directly from the unconstrained linear-least-squares solutions for the parameters $\langle I_{\lambda}/\Sigma_{12}\rangle$ and $b_{\lambda}$, this procedure places higher weights on those VIMOS fibers that fall on higher density portions of the cluster.   In essence, we used our knowledge of the cluster density profile to extract only the information that interests us, namely, that part of the emission that is correlated with the cluster's mass density profile. At some wavelengths, the best fit is $\langle I_{\lambda}/\Sigma_{12}\rangle=0$ with very low noise.  We verified that these wavelengths coincide with those at which a naive linear fit yielded negative values for $\langle I_{\lambda}/\Sigma_{12}\rangle$. Thus there is no evidence for density correlated emission at these wavelengths. At these wavelengths, the emission due to axion decay is bounded from above by the brightness of the sky background, and so we used $\langle I_{\lambda}/\Sigma_{12}\rangle\leq b_{\lambda}/\langle\Sigma_{12}\rangle$ to obtain a conservative upper limit on the flux. Decaying axions will produce line emission, so it might seem that an additional continuum subtraction might be in order. The continuum component of the sky background, however, is already subtracted using the techniques discussed, and an additional continuum subtraction step would be erroneously aggressive.

As a test of our sky-subtraction technique, we also reimplemented the sky-subtraction technique of Ref.  \cite{Bershady:1990sw} and implemented an `on-off' subtraction by defining fibers further than $23''$ (A2667) or $72''$ (A2390) from the cluster center (defined by the highest density point in the density maps) as `sky' fibers, spatially averaging the flux of these sky fibers at each wavelength, and subtracting the resulting sky spectrum from each pixel in the `on' cluster region. In this case, sky emission was directly estimated from the data rather than modeled. In this case, the best fit for the signal is given by 
\begin{equation}
\left \langle\frac{I_{\lambda}}{\Sigma_{12}}\right\rangle=\frac{\sum_{i}\frac{I_{\lambda,i}\Sigma_{12,i}}{\sigma_{\lambda,i}^{2}}}{\sum_{i}\frac{\Sigma_{12,i}^{2}}{\sigma_{\lambda,i}^{2}}},\label{bf}
\end{equation}
where $i$ is a label for the density at the location of a given IFU fiber, and $\sigma_{\lambda,i}$ is the error in the specific intensity.

In the case of A2667, even the fibers furthest from the cluster center fall on portions of the cluster where emission due to axion decays will be of the same order of magnitude as at the center. The sky-subtraction technique of Ref. \cite{Bershady:1990sw} is thus entirely inappropriate for our data on A2667, as it will subtract out a substantial fraction of any signal and return unjustifiably stringent limits to emission from axion decay.

\begin{figure}
\includegraphics[width=8 cm]{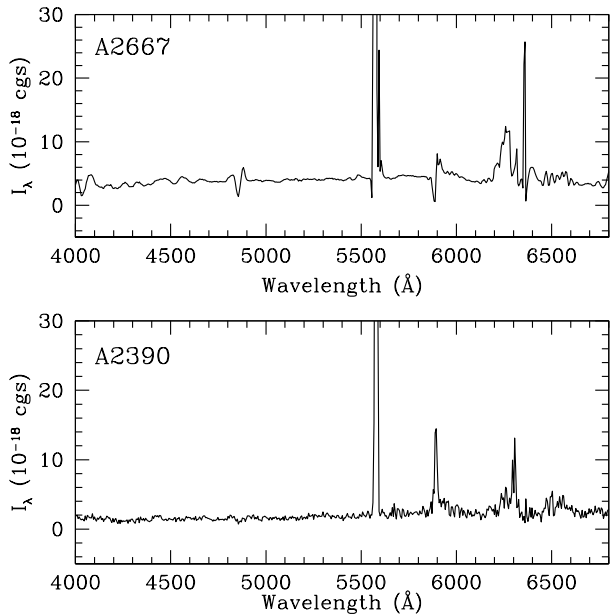}
\caption{Average one-dimensional sky subtracted spectra of clusters A2667 and A2390. Intensity is in units of $10^{-18}~{\rm ergs}~{\rm cm}^{-2}~{\rm s}^{-1}~{{\rm \AA}}^{-1}~\mbox{arcsec}^{-2}$. Poorly subtracted sky emission lines at $5577\rm{\AA}$, $5894\rm{\AA}$, and $6300\rm{\AA}$ have not been removed.}
\label{onoff}
\end{figure}

\begin{figure}
\includegraphics[width= 8 cm]{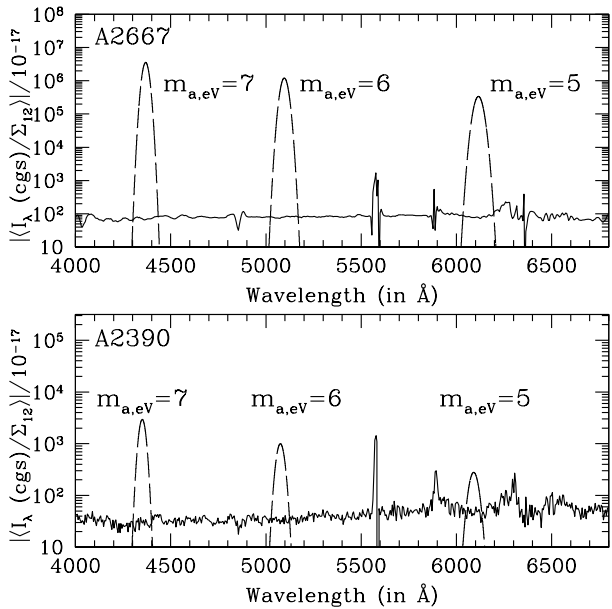}
\caption{Constraints on $\langle I_{\lambda}/\Sigma_{12}\rangle$ as a function of wavelength $\lambda$ for A2667 and A2390. CGS units for specific intensity are ${\rm ergs}~{\rm
cm}^{-2}~{\rm s}^{-1}~{{\rm \AA}}^{-1}~\mbox{arcsec}^{-2}$, and $\Sigma_{12}=\Sigma/(10^{12} M_{\odot}~\rm{pix}^{-2})$, where $\Sigma$ is the projected mass density of the cluster, measured using strong lensing. The over-plotted dashed lines are theoretical Gaussian spectra for axion decays, with central wavelength $\lambda_{0}$,  corresponding to an axion mass of $m_{\rm{a},\rm{eV}}=24,800{{\rm \AA}}(1+z)/\lambda_{0}$. The predicted amplitude is set by Eq.~(\ref{lintens}), and exceeds the measured values in both the top panel ($\xi=1.0$) and the bottom panel ($\xi=0.03$).}
\label{fluxlimit}
\end{figure}

For A2390, the effective field of view is much larger, and so the emission expected from axion decays in the outer fibers is much less. Over most of the wavelength range of our data for A2390, the different sky-subtraction techniques agreed to within a factor of two, leading us to believe that our sky-subtraction technique is trustworthy. We used the value for $\langle I_{\lambda}/\Sigma_{12}\rangle$ obtained using our sky-subtraction technique, as it is desirable to use the same sky-subtraction method for both clusters to be self-consistent. Equation (\ref{bf}) and the corresponding best-fit result in the constrained case essentially yield one-dimensional cluster spectra, rescaled by the cluster density, as shown in Figs.~\ref{onoff} and \ref{fluxlimit}. The specific intensity values in Fig.~\ref{onoff} were obtained by multiplying the best-fit values of $\langle I_{\lambda}/\Sigma_{12}\rangle $ from Eq.~(\ref{bf}) by the mean $\langle \Sigma_{12}\rangle= \left(\sum_{i} \Sigma_{12,i}/\sigma_{\lambda,i}^{2}\right)/\left(\sum_{i}1/\sigma_{\lambda,i}^{2}\right)$. The plotted spectrum is thus not the best-fit spectrum at any particular fiber, but an average cluster spectrum. The signal to noise ratio of the one-dimensional spectrum appears to be higher for A2667 than for A2390, in spite of the lower effective fiber number of the data cube for A2667. We believe that this is the case because the data cube for A2390 was built using four nights of data, with slight variations in sky intensity and efficiency from night to night. The subtraction is poorest around the prominent sky line at $5577\rm{\AA}$. There is no obvious candidate for an emission line due to axion decay. 

\subsection{Limits on the two-photon coupling of axions}
The expected strength of an axion decay line is set by $m_{\rm{a},\rm{eV}}$ through Eq.~(\ref{lintens}), and  the expected Gaussian line profiles are shown on top of our appropriately normalized upper limits to flux in Fig.~\ref{fluxlimit} for several candidate axion masses. The narrow feature at $5577\rm{\AA}$, present in both panels of Fig.~\ref{fluxlimit}, arises from the imperfect subtraction of a sky emission line. In the absence of a candidate axion decay line, we proceed to put an upper limit on the coupling strength $\xi$ of an axion to two photons.

Since our best-fit values for $R_{\lambda}\equiv\langle I_{\lambda}/\Sigma_{12}\rangle$ at each wavelength come with an error estimate $\sigma_{\lambda}$, we can calculate a $95\%$-confidence limit to the line flux. We assume that the distribution of noise peaks is Gaussian, and so the probability that an axion decay associated with a particular value of $R_{\rm{a},\lambda}$ yields a measured best-fit value less than $R_{\lambda}$ is
\begin{equation}
P_{\lambda}=\frac{1}{\sqrt{2\pi} \sigma_{\lambda}}\int_{-\infty}^{R_{\lambda}-R_{\rm{a},\lambda}} e^{\frac{-x^{2}}{2\sigma_{\lambda}^{2}}} dx.\label{probeq}
\end{equation}
Eq.~(\ref{probeq}) yields the $95\%$-confidence limit on intensity from axion decay:\begin{equation}
R_{\rm{a},\lambda}\leq R_{\lambda}+1.65\sigma_{\lambda}.\label{uplim1}\end{equation}
At those wavelengths where the best-fit value is
$R_{\lambda}=0$, we have taken the roughly homogeneous intensity
of the sky as a very conservative upper limit on the
intracluster emission. This is many $\sigma$ above the
$95\%$-confidence limit, and so at these wavelengths, we just
take $R_{\rm{a},\lambda}\leq R_{\lambda}$ without making our
estimate of the upper limit too conservative. Ultimately, we
wish to combine the upper limits to flux from the two
clusters. One of the advantages of working with two clusters at
slightly different redshifts is that rest-frame wavelengths
falling near sky lines (where limits to flux are generally quite
poor) at one redshift may no longer fall on sky emission lines
at the redshift of the second cluster. When this is the case, we
excise wavelengths falling on or near sky lines from each
spectrum. To account for all the flux in a given candidate axion
line, in each cluster spectrum, we calculate the average
intensity of non-excised data points in a
$24,800\left[(1+z)\sigma/c\right]~{\rm \AA}~m_{{\rm a},{\rm
eV}}$ window around a series of putative line centers spanning
the probed axion mass range. We weight the noise in the usual
way. Assuming that our spectra uniformly sample this bin and
that flux errors are uniform across the bin, we see by
integrating the Gaussian profile given in Eq.~(\ref{lintens})
that
\begin{equation}
     R_{\rm{a},\lambda}=\frac{2.30\times 10^{-18}~ \xi^{2}m_{{\rm
     a},{\rm eV}}^{7}}{(1+z_{{\rm cl}})^{4}S^{2}(z_{{\rm
     cl}})\sigma_{1000}}~{\rm cgs}.
\end{equation}
If axions have the standard thermal-freezeout abundance
[Eq.~(\ref{thermalfreeze})], then the limit on the axion
coupling is given by
\begin{equation}
\begin{array}{c}
     \xi \leq
     \left[\frac{\sigma_{1000}(1+z_{\rm{cl}})^{4}S^{2}(z_{\rm{cl}})m^{-7}_{\rm{a},\rm{eV}}(\lambda)R_{\rm{a},\lambda}}{2.30\times10^{-18}~\rm{cgs}}\right]^{1/2}
\end{array}. 
\label{fluxxi}
\end{equation}
If the cosmological axion abundance takes on some other value
$\Omega_{\rm a}h^2$, then the limit becomes
\begin{equation}
\begin{array}{c}
     \xi \sqrt{\Omega_{{\rm a}}h^{2}} \leq
     \left[\frac{\sigma_{1000}(1+z_{\rm{cl}})^{4}S ^{2}
     (z_{\rm{cl}})m^{-6}_{\rm{a},\rm{eV}}(\lambda)
     R_{\rm{a},\lambda}}{3.48\times10^{-16}~\rm{cgs}}\right]^{1/2}.
     \end{array} 
\label{fluxxiomega}
\end{equation}
Since our real bins are not uniformly sampled (because of the
excision of wavelengths that fall on sky emission lines) and
since the errors scale with the intensity value at a given
wavelength, we make a small correction to this expected
value. Specifically, 
\begin{eqnarray}
\nonumber
R_{{\rm a},\lambda}=\frac{2.68\times 10^{-18}~\xi^{2} m_{{\rm a},{\rm eV}}^{7}}{(1+z_{{\rm cl}})^{4}S^{2}(z_{{\rm cl}})}\frac{\sum_{j\in T_{\lambda}}\frac{G_{j}}{\sigma_{j}^{2}}}{\sum_{j\in T_{\lambda}}\frac{1}{\sigma_{j}^{2}}}~{\rm cgs},\\
\mbox{where}~G_{j}=e^{-\frac{(\lambda_{j}/(1+z_{{\rm cl}})-\lambda_{{\rm a}})^{2}c^{2}}{2\sigma^{2}\lambda_{{\rm a}}^{2}}},\label{alim}
\end{eqnarray} $T_{\lambda}$ is the set of all non-excised wavelengths lying within the bin centered at wavelength $\lambda$, and $j$ labels wavelengths.
The quality of sky background subtraction may vary as a result
of spatial and temporal variations in the sky background from
night to night. If it is not due to axion decay, the density
correlated emission might also genuinely vary between
clusters. The quantity $R_{\rm{a},\lambda}$, however, will by
definition be independent of these factors. A simple error
weighted mean of the upper limits obtained from the two clusters
would thus erroneously increase the upper limit placed on
$\xi$. If two clusters yield different best-fit values for
$R_{\lambda}$, $R_{\rm{a},\lambda}$ must be bounded from above
by the lesser of these two. By comparing upper limits to
$R_{\lambda}$ obtained from A2390 with those obtained from A2667
and choosing the lowest value at each wavelength, we obtained
the maximum values of $R_{\rm{a},\lambda}$ consistent with the
data. We then applied Eq.~(\ref{alim}) to obtain an upper
limit on $\xi$ consistent with the spectra of both clusters. To
account for variation in the upper limits to $\xi$ arising from
systematic errors in the cluster mass profiles, we repeated the
preceding analysis,  drawing $\Sigma_{12,i}$ from the best-fit
NFW (Navarro, Frenk, and White) and King profiles to the cluster
mass profiles.

\begin{figure}
\includegraphics[width= 8 cm]{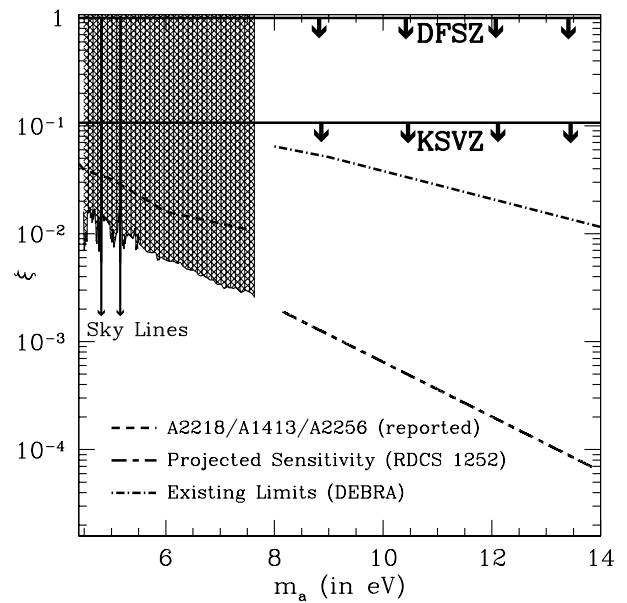}
\caption{Comparison of existing limits to the two-photon coupling of a $4.5~\rm{eV}-14~\rm{eV}$ axion with the projected sensitivity of our proposed observations of lensing cluster RDCS 1252 ($z=1.237$). Flux limits and density profiles were assumed to be the same as those of A2667/A2390. The best existing upper limits to $\xi$ in the higher mass window come from limits to the Diffuse Extragalactic Background Radiation (DEBRA), and were rescaled for consistency with today's best-fit $\Lambda$CDM parameters and recent measurements \cite{debra,debra2}. The limits reported in this and previous work, 
derived using optical spectroscopy  of galaxy clusters, are shown for comparison \cite{ressellt,Bershady:1990sw}. Regions inaccessible due to night-sky emission lines are marked with narrow black arrows. The two solid horizontal lines indicate the predictions of the DFSZ and KSVZ axion models; the downward arrows indicate that $\xi$ is \textit{theoretically} uncertain.}
\label{fig: project}
\end{figure}

Analytic expressions for the volumetric and surface mass density
for NFW and King profiles are reviewed in Appendix
\ref{twob}. We determined the mass profile parameters ($a$
and $\sigma$ for King profiles, $c$ and $\sigma$ for NFW
profiles) by fitting to our strong-lensing density maps. Using
these different density profiles and assuming that the mass
fraction in axions is
$x_{\rm{a}}=\Omega_{\rm{a}}h^{2}/(\Omega_{{\rm m}}h^{2})$, we
obtain limits to $\xi$. The cluster density at the location of a
given IFU fiber varies from profile to profile, and so different
fibers receive higher weights when a one-dimensional spectrum is
extracted. This explains the variation in upper limits to $\xi$
that arises when different density profiles are assumed. We show
the most conservative (with respect to choice of density
profile) limit on $\xi$ (assuming the thermal-freezeout
abundance of axions) at each candidate axion mass in
Fig.~\ref{fluxxilimitbin}. The upper limits to $\xi$ in adjacent
points along the $m_{\rm{a},\rm{eV}}$ axis are correlated due to
overlapping bins. The narrow black arrows near
$m_{\rm{a},\rm{eV}}=5.43$ and $4.83$ mark sharp night-sky lines
at $\lambda=5577\rm{\AA}$ and $\lambda=6300{\rm \AA}$, where sky
subtraction is unreliable and useful limits to $\xi$ cannot be
obtained. Limits on $\xi$ and specific intensity at the putative
line center are displayed for several candidate masses in Table
\ref{limtable1}. Our data rule out the canonical DFSZ and KSVZ
axion models in the mass window $4.5\leq m_{\rm{a},\rm{eV}}\leq
7.7$, as seen in Fig.~\ref{fig: project}. However, theoretical
uncertainties motivate the search for axions with values of
$\xi$ smaller than those allowed by the canonical KSVZ and DFSZ
models.

\begin{figure}
\includegraphics[width= 8 cm]{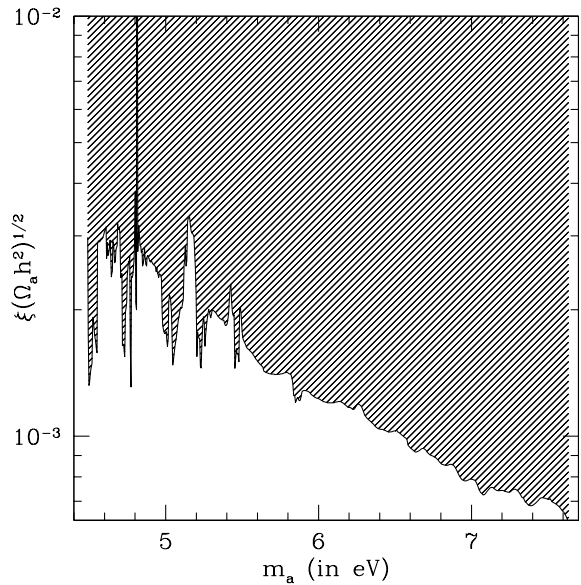}
\caption{Limits on the combination $\xi\left(\Omega_{{\rm a}}h^{2}\right)^{1/2}$, derived directly from upper limits to the intracluster flux of A2667 and A2390. Our data exclude the shaded region. Data analysis proceeds as in the thermal case, but appropriate (more general) expressions for the intensity $I_{\lambda}$ are used. These constraints do not depend on the assumption that axions are produced thermally at early times.}
\label{fluxcombolimitbin}
\end{figure}

If we relax the assumption that the cosmological axion abundance
be given by Eq.~(\ref{thermalfreeze}), then our null search
implies the bound, shown in Fig.~\ref{fluxcombolimitbin}, to the
combination $\xi (\Omega_{\rm a} h^2)^{1/2}$.  We see that if
$\xi\sim 10^{-1}$, as is the case in the KSVZ model, then our results imply an upper limit $\Omega_{\rm a}h^2 \lsim 10^{-4}$ in our mass range, roughly two orders of
magnitude stronger than CMB/LSS limits \cite{hanraffelt}, which
probe densities down to $\Omega_{\rm a} h^2\sim10^{-2}$.

\subsection{Revision of past telescope constraints to axions}
\begin{figure}
\includegraphics[width= 8 cm]{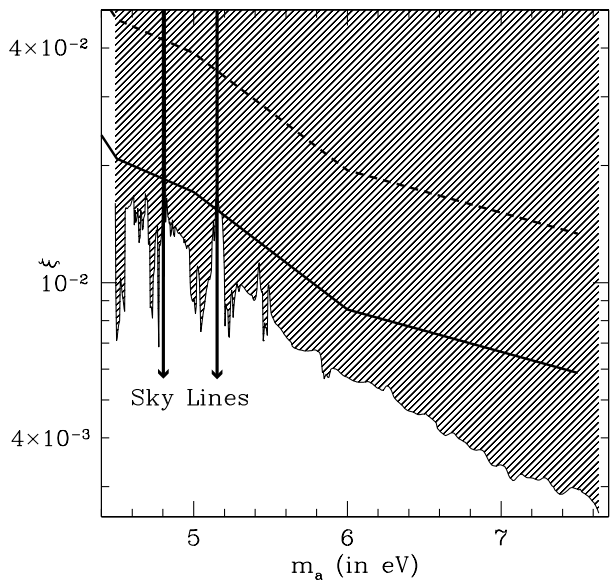}
\caption{Upper limits to the two-photon coupling parameter $\xi$ of the axion, derived directly from upper limits to the intracluster flux of A2667 and A2390. Our data exclude the shaded region. The solid and dashed lines show the upper limits reported in Refs. \cite{ressellt,Bershady:1990sw}, adjusted (optimistically and pessimistically) for differences between today's best-fit measurements of the cosmological parameters/cluster mass profiles and the assumptions in Refs. \cite{ressellt,Bershady:1990sw}. Details are discussed in Appendix \ref{assume}. The mass range $4.5~{\rm eV}\leq m_{{\rm a}}\leq 7.7~{\rm eV}$ arises from the $4000\rm{\AA}$--$6800\rm{\AA}$ usable wavelength range of VIMOS, which is smaller than that of the KPNO spectrograph used in Ref. \cite{Bershady:1990sw}. The narrow black arrows near $5.43~{\rm eV}$ and $4.83~{\rm eV} $ mark the sharp night-sky lines at $5577\rm{\AA}$ and 
$6300{\rm \AA}$, where sky subtraction is unreliable and useful limits to $\xi$ cannot be obtained. The shaded exclusion region is derived by applying the cluster density profile (strong-lensing map, best-fit NFW profile, or best-fit King profile) at each candidate axion mass that yields the most conservative upper limit on $\xi$.}
\label{fluxxilimitbin}
\end{figure}

As can be seen from Eq.~(\ref{lintens}), and from the fact that the mass fraction of the cluster in axions is $x_{\rm{a}}=\Omega_{\rm{a}}h^{2}/(\Omega_{{\rm m}}h^{2})$, the upper limit on $\xi$ derived from a given upper limit on flux depends on the cluster mass model used and the cosmological parameters assumed. References \cite{Bershady:1990sw,ressellt} date to a time when the observationally favored cosmology was sCDM (Standard Cold Dark Matter: $h=0.5$, $\Omega_{{\rm m}}=1.0$, $\Omega_{\Lambda}=0$). Moreover, the King profiles assumed in those analyses of A2218, A2256, and A1413 were based on available x-ray emission profiles of the chosen clusters \cite{ressellt,Bershady:1990sw} (and references therein).  The advent of modern x-ray instruments has improved x-ray-derived cluster mass profiles, and gravitational-lensing studies have allowed measurements of cluster mass profiles, free of the dynamical assumptions required to obtain a density profile from an x-ray temperature map. The quoted upper limits of Refs. \cite{ressellt,Bershady:1990sw} must thus be rescaled, and we have done so up to an ambiguity in slit placement for A2218; details are discussed in Appendix \ref{assume} and the rescaled limits from past work are shown alongside our own in Fig.~\ref{fluxxilimitbin}. Our limits improve on the rescaled limits of Ref. \cite{ressellt,Bershady:1990sw} by a factor of $2.1-7.1$. Our final measurement of $\langle I_{\lambda}/\Sigma_{12}\rangle$ is only noise limited at a small fraction ($\simeq 10\%$) of the available wavelength range. The dominant uncertainty is systematic and comes from sky subtraction. The expected improvement estimated in the introduction assumed that we are limited by Poisson noise in the measured flux, and is thus naive.

\begin{table}
\caption{Upper limits to central line intensity and $\xi$ at several candidate axion masses, derived directly from sky subtracted spectra of A2667 and A2390.}
\begin{ruledtabular}
\begin{tabular}{rrr}
$m_{\rm{a},\rm{eV}}$&$\left\langle I_{\lambda}/\Sigma_{12}\right\rangle\left\langle\Sigma_{12}\right\rangle$ (cgs) &$\xi$\\ \hline
$ 4.5 $ &$  1.83\times 10^{-19} $ & $ 7.17\times 10^{-3}$  \\ \hline 
$ 5 $ &$  6.04\times 10^{-19} $ & $ 9.00\times 10^{-3}$  \\ \hline 
$ 6 $ &$  8.74\times 10^{-19} $ & $ 5.72\times 10^{-3}$  \\ \hline 
$ 6.5 $ &$  9.91\times 10^{-19} $ & $ 4.60\times 10^{-3}$  \\ \hline 
$ 7 $ &$  9.13\times 10^{-19} $ & $ 3.41\times 10^{-3}$  \\ \hline 
$ 7.5 $ &$  1.11\times 10^{-18} $ & $ 2.95\times 10^{-3}$  \\ \hline 
$ 7.65 $ &$  8.96\times 10^{-19} $ & $ 2.47\times 10^{-3}$ \\
\end{tabular}
\end{ruledtabular}
\label{limtable1}
\end{table}


Previous cluster searches for axions used long-slit spectroscopy. Our use of IFU data is novel, and it is conceivable that peculiarities of the data-reduction techniques used in IFU spectroscopy may affect the sensitivity of our search. To explore this possibility, we have conducted a simulation.

\subsection{Simulation of Analysis Technique}
\begin{figure}
\includegraphics[width=80mm]{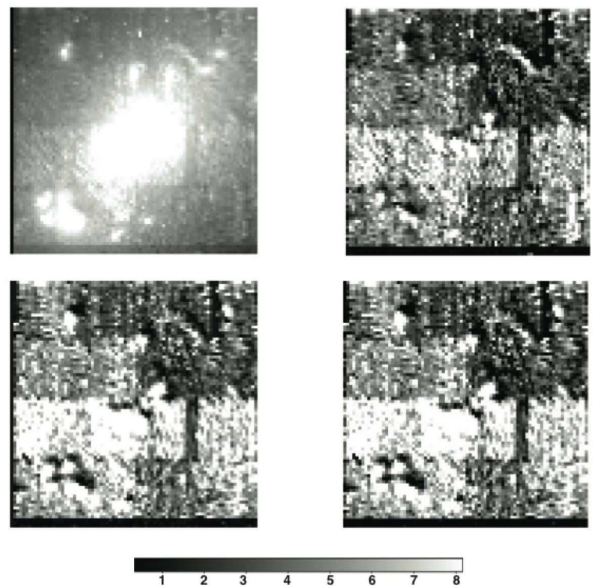}
\caption{The upper left panel of this figure shows a simulated $4255.2\rm{\AA}$ slice of the A2667 IFU data cube, with an axion-decay emission line inserted corresponding to $m_{\rm{a},\rm{eV}}=7.2$ and $\xi=0.011$. The flux scale is in units of $10^{-18}~\rm{erg}~\rm{s}^{-1}~\rm{cm}^{-2}~\rm{\AA}^{-1}$. This slice, which lies at the expected line center, shows evidence of the inserted axion line. The resulting `emission' clearly traces the cluster mass density profile. The lower left panel of this figure shows a simulated slice of the same data cube, but at $5267.2\rm{\AA}$, well away from the line center. As expected, no signature of axion emission is present this far away in wavelength from the line center. The upper/lower right panels of this figure show $4255.2\rm{\AA}$/ $5267.2\rm{\AA}$ slices, respectively, of the actual A2667 IFU data cube used for our analysis.}
\label{simpic}
\end{figure}

We simulate axion-decay emission in our data cube for A2667, using Eq.~(\ref{lintens}) and our lensing derived projected density maps. We did this at a range of $10$ candidate axion masses spanning the full mass range of our search. We used $3$ or $4$ different values of $\xi$ at each candidate mass. The first value was chosen to be slightly below ($5-10\%$) the limit on $\xi$ set by preceding techniques, while the second was chosen to be slightly above the upper limit. The third and fourth values were chosen to be in considerable (factors of $2$ and $10$, respectively) excess of the upper limit. For all simulated axion masses, visual inspection of the data cube yields clear evidence for the inserted line when $\xi$ exceeds the imposed upper limit. An example is shown in Fig.~\ref{simpic}. After inspecting the data cubes visually, we applied the routines used for the preceding analysis to produce one-dimensional spectra for each cube. We then applied the same routine used to extract upper limits to $\xi$ to recover the simulated $\xi$ value. When the simulated value of $\xi$ exceeded the upper limit, we recovered the correct answer in all cases to a precision of $5-10\%$. This leads us to believe that our technique is robust and our upper limits reliable. References \cite{ressellt,Bershady:1990sw} supplement upper limits to $\xi$ derived directly from flux limits with limits obtained from a cross-correlation analysis. We do the same, using our data on A2667 and A2390.

\subsection{Cross-Correlation Analysis}
\begin{figure}
\includegraphics[width= 8 cm]{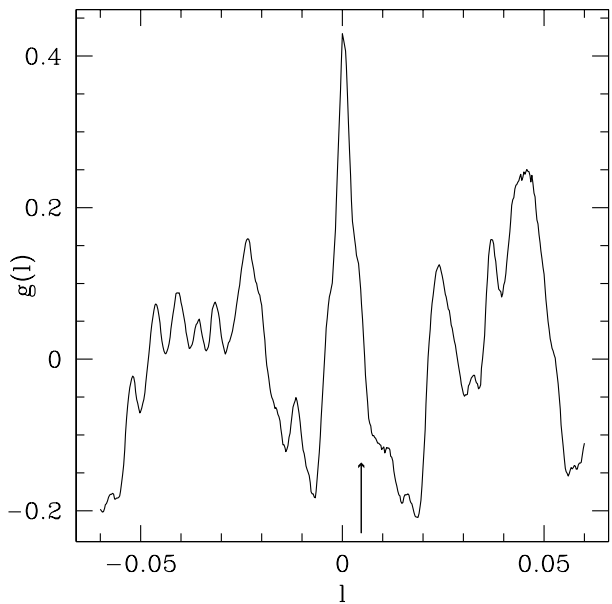}
\caption{Cross-correlation function $g(l)$ between sky subtracted spectra of clusters A2667 and A2390. No peaks with the desired statistical significance were seen, at the desired lag or elsewhere. A single cross-correlation peak is near the expected lag $l=0.00498$ for an intracluster emission line common to A2667 and A2390. However, it is not statistically significant.\label{fig: crosscorractual}}
\end{figure}

If there is an emission line at the same wavelength in the rest frame of both clusters, the function 
\begin{equation}
g(l)=\frac{\int I_{1}(x)I_{2}(x+l) dx}{\left[\int I_{1}^{2}(x) dx \int I_{2}^{2}(x) dx\right]^{1/2}},
\end{equation}
will have a peak at the lag $l_{0}=\ln{\left[\left(1+z_{{\rm A}2667}\right)/\left(1+z_{{\rm A}2390}\right)\right]}$, where $x=\ln{\lambda}$, $I_{1}(x)$ and $I_{2}(x)$ are the specific intensities of galaxy clusters A2667 and A2390, and $z_{{\rm A}2667}$ and $z_{{\rm A}2390}$ are their redshifts \cite{ressellt,Bershady:1990sw}. A statistically significant peak in $g(l)$ would indicate the existence of an intracluster emission line at unknown wavelength (and correspondingly unknown axion mass), which could then be searched for more carefully in the individual spectra.
Peaks due to noise may arise either due to the roughly Gaussian fluctuations in flux of the individual spectra, or due to imperfectly subtracted flux around sharp sky emission lines. It is thus appropriate to mask out bright sky lines. If we assume that the distribution of remaining noise peaks is Gaussian, then the probability that a cross-correlation peak with height greater than or equal to $s$ is due to noise is \cite{ressellt,Bershady:1990sw}
\begin{equation}
P(\geq s)=\int_{s}^{\infty} \frac{e^{-x^{2}/(4\sigma_{g}^{2})}dx}{\sqrt{\pi}\sigma_{g}}=1-\mbox{Erf}\left(\frac{s}{2\sigma_{g}}\right).\label{ccsig}
\end{equation}
Here, $\sigma_{g}$ is the rms value of the antisymmetric component of $g(l)$ and provides an estimate of the correlation due to noise, since a Gaussian signal leads to a symmetric correlation function \cite{ressellt,td}. Equation (\ref{ccsig}) determines the statistical significance of peaks in $g(l)$. Our analysis of correlated spectra follows the treatment of Ref. \cite{td}. We calculate $g(l)$ using the sky subtracted one-dimensional spectra of A2667 and A2390. With a cross-correlation technique, we are able to perform a blind search for cluster rest-frame emission. We find no statistically significant ($>2\sigma_{g}$) cross-correlation peaks, as shown in Fig.~\ref{fig: crosscorractual}. 

\begin{table}
\caption{Upper limits to $\xi$ at several candidate axion masses, obtained from a simulation of the cross-correlation method, using spectra of A2667 and A2390.\label{cct}}
\begin{ruledtabular}
\begin{tabular}{cc}
$m_{\rm{a},\rm{eV}}$&$\xi$\\ \hline
$ 4.5 $ &$ 2.73\times 10^{-2}$ \\ \hline 
$ 5 $ &$ 1.08\times 10^{-2}$ \\ \hline 
$ 6.0 $ &$ 9.35\times 10^{-3}$ \\ \hline 
$ 6.5 $ &$ 6.90\times 10^{-3}$ \\ \hline 
$ 7 $ &$ 4.44\times 10^{-3}$ \\ \hline 
$ 7.5 $ &$ 4.31\times 10^{-3}$ \\ \hline 
$ 7.65 $ &$ 2.16\times 10^{-2}$ \\ 
\end{tabular}
\end{ruledtabular}
\end{table}

We simulate our cross-correlation based search for an intracluster line in order to set alternative upper limits to $\xi$ \cite{ressellt,Bershady:1990sw}. The limits reported in Ref. \cite{Bershady:1990sw} were obtained using this technique. In our simulation, we introduced `fake' axion decay lines into both spectra and calculated the resulting $g(l)$ for a variety of values $m_{\rm{a},\rm{eV}}$, thus simulating the cross-correlation search for evidence of intracluster emission. $\xi$ was initially set to a value for which a very significant peak in $g(l)$ appeared, and then ramped down until the peak at $l_{0}$ ceased to be statistically significant; that is, until $P(s)>5\%$. At that point, an upper limit on $\xi$ was set. The cross-correlation peaks due to the axion line are well fit by Gaussian curves throughout the $m_{\rm{a},\rm{eV}}-\xi$ parameter space. Since the cross-correlation function includes the contribution of flux away from the line center, rebinning was unnecessary. 

To further distinguish between signal and noise peaks, we follow Ref. \cite{ressellt} in using the usual criterion, $|l-l_{0}|\leq \sigma$, where $\sigma$ is the width of the best Gaussian fit to $g(l)$ around a cross-correlation peak. We use this criterion in both the simulation and the cross-correlation search\cite{ressellt}. The resulting limits to $\xi$ at a series of candidate axion masses are shown in Table \ref{cct}, and are on average a factor of $\sim 1.5$ less stringent than those derived directly from flux.

One aspect of the correlation analysis of Ref. \cite{ressellt} is troubling. Two noisy, imperfectly sky subtracted spectra were correlated to search for a signal. The analysis of Ref. \cite{ressellt}, however, uses one real spectrum (containing noise and an imperfectly subtracted sky-background signal) with an artificial axion line inserted, and a second, noiseless, template spectrum, containing only the artificial axion line, but no imperfectly subtracted sky component. Thus, the method simulated in Ref. \cite{ressellt} is not the same as the method used to actually search for evidence of an intracluster line, and by artificially reducing the noise budget of the simulation, could lead to artificially stringent constraints. The appropriate way to simulate the cross-correlation analysis is to correlate two real spectra with artificial axion lines inserted, as we have done. Our data also place limits on the decay of other relics.

\subsection{Sterile neutrinos}

Our data might also be used to constrain the decay rate of other
$\sim5~{\rm eV}$ relics, such as sterile neutrinos
\cite{dodwid,dolgovhansen,shifuller,shaposh}.  Although the
prevailing paradigm places the sterile-neutrino mass in the keV
range, some experimental data can be fit by introducing a
hierarchy of sterile neutrinos, at least one of which is in the
$1-10~{\rm eV}$ range and could oscillate to produce photons in
our observation window \cite{gouvea1,gouvea2}. In our notation
and in the $m_{e^{-},\mu,\tau}\ll m_{s}$ limit (where $m_s$ is
the sterile-neutrino mass), the intensity of
this signal is
\begin{widetext}
\begin{equation}
     \left\langle\frac{I_{\lambda,{\rm
     s}}}{\Sigma_{12}}\right\rangle=2.4\times 10^{-18}
     \frac{B m_{{s},{\rm
     eV}}^{8}\exp{\left[-\left(\lambda_{r}-\lambda_{{\rm
     s}}\right)^{2}c^{2}/\left(2\lambda_{{\rm
     s}}^{2}\sigma^{2}\right)\right]}}{\sigma_{1000} (1+z_{{\rm
     cl}})^{4}S^{2}(z_{{\rm cl}})} {\rm cgs},
\end{equation}
\end{widetext} 
where $B$ is a model-dependent normalization factor, the
oscillation is parameterized by a cumulative mixing angle
$\theta$, and the additional power of mass arises from  the
late-time abundance of sterile neutrinos \cite{abfullerpatel}:
\begin{equation}
     \Omega_{{\rm s}}h^{2}=0.3\times m_{ s,{\rm
     eV}}^{2}\sin^{2}2\theta.
\end{equation}

The flux limits in Table \ref{limtable1} impose the constraints
$B\leq 8.03\times 10^{-5}, ~1.14\times 10^{-4}, ~9.41\times
10^{-5}, ~3.82\times 10^{-5}, ~2.28\times 10^{-5},
1.16\times10^{-5}, ~8.12\times 10^{-6}$, and $5.61\times
10^{-6}$ for sterile-neutrino masses of $m_{s,{\rm
eV}}=4.50,~5.00,~5.50,~6.00,~6.50,~7.00,~7.50$, and $7.65$,
respectively. In conventional models,
$B=\sin^{4}{\left(2\theta\right)}/10^{11}$. The parameter $B$ encodes the
effects of both the early-universe production and the decay of
sterile neutrinos, which occurs at the rate
$\Gamma_{s\to\nu+\gamma}=6.8 \times10^{-38}{s}^{-1}m_{s,{\rm
eV}}^{5}\sin^{2}{2\theta}$
\cite{wolfpal,glashow,abfullertucker,barger2}. By
definition, $B\leq10^{-11}$, and so optical data only constrain
sterile neutrinos if some novel mechanism increases the
oscillation rate $\Gamma_{s\to\nu+\gamma}$ by many orders of
magnitude. The sharp disparity between x-ray and optical
constraints results from the $\Gamma\propto m_{s}^{5}$ scaling
of the decay rate.

\subsection{Future Work}
We have demonstrated the utility of applying integral field spectroscopy in concert with lensing data to search for axions in $z\simeq 0.2$ galaxy clusters. Our technique could also be profitably applied to higher redshift galaxy clusters. Although flux falls off as $I_{\lambda}\propto (1+z_{{\rm cl}})^{-4}$, the fact that we are pushing to a higher mass range $m_{{\rm a},{\rm eV}}=24,800{\rm \AA}\left(1+z_{{\rm cl}}\right)/\lambda_{{\rm a}}$ increases the expected signal. Since $I_{\lambda}\propto m_{{\rm a},{\rm eV}}^{7}$, the expected signal actually increases as $I_{\lambda}\propto \left(1+z_{{\rm cl}}\right)^{3}$. The most distant known lensing cluster is RDCS 1252, at redshift $z=1.237$ \cite{rosati}.

Using existing weak-lensing mass maps for this cluster \cite{lombardi} and creating our own strong-lensing maps, we should be able to obtain a sky subtracted, spatially weighted spectrum of this cluster, attaining flux levels similar to those we have obtained for A2667 and A2390. We will thus be able to search for emission from decaying axions in the mass window $8~{\rm eV}\leq m_{{\rm a}} \leq14~\rm{eV}$. Assuming identical cluster density profiles and flux limits, we estimate the range of $\xi$ values accessible with a telescope search for cluster axions in RDCS 1252. The tightest existing constraints to decaying relic axions in this mass window come from limits to the diffuse extragalactic background radiation (DEBRA) \cite{debra,debra2}. As shown in Fig.~\ref{fig: project}, a VIMOS IFU search for axions in this mass window would detect very weakly coupled axions, or alternatively, improve upper limits to $\xi$ by two orders of magnitude. Applying Eq.~(\ref{thermalfreeze}), we see that $8~\rm{eV}-14~\rm{eV}$ axions would freeze out with abundance $0.12\leq\Omega_{{\rm m}} \leq 0.21$. An axion detection in this mass window could thus account for most of the dark matter; a telescope search in this mass window would provide a useful check of LSS constraints to axion properties. Future discoveries of even higher redshift clusters could allow heavier axion mass windows to be probed with cluster observations. 

\section{Conclusions}
The axion hypothesis offers attractive solutions to both the strong CP and dark-matter problems. A series of null searches and astrophysical constraints has narrowed down the parameter space of the axion to two mass windows, one between $10^{-5}~\rm{eV}$ and $10^{-3}~\rm{eV}$, and the other between $3~\rm{eV}$ and $20~\rm{eV}$. Previous searches for optical emission from decaying axions in galaxy clusters have constrained the two-photon coupling of the axion in the latter window. We have searched for axion-decay light in the galaxy clusters A2667 and A2390, taking advantage of strong-lensing mass maps of A2667/A2390 to free our analysis of dynamical assumptions. Use of the VIMOS IFU allowed an increase in effective collecting area, thus increasing the sensitivity of our axion search. We observed no evidence for emission from decaying axions in the mass window between $4.5~\rm{eV}$ and $7.7~\rm{eV}$. 

Conservatively, we improve on constraints to the two-photon coupling $\xi$ of axions by a factor of $\simeq 3$, averaged over the entire mass range we explore. This work presents the first application of IFU spectroscopy to constrain the nature of the dark matter and not just its spatial distribution. To check that the stringency of our constraints is not an artifact of the rather complicated data-reduction techniques inherent to IFU spectroscopy, we have simulated our technique by introducing fake axion lines into our data cubes. Our analysis technique accurately recovers the value of $\xi$, and the axion's signature fades into the sky-background as $\xi$ is ramped down below our reported upper limits. Our simulations demonstrate the robustness of our technique, and our work highlights the potential of IFU spectroscopy for more sensitive exploration of the axion mass window between $8~\rm{eV}$ and $14~\rm{eV}$.

\begin{acknowledgments}
The authors thank Ted Ressell and Matthew Bershady for helpful discussions. D.G. was supported by a Gordon and Betty Moore Fellowship. G.C. acknowledges support from the European Community via the Marie Curie European Re-Integration Grant n.029159. J-P.K. acknowledges support from the CNRS. M.K. was supported in part by DoE DE-FG03-92-ER40701, NASA NNG05GF69G, and the Gordon and Betty Moore Foundation. A.W.B thanks the Alfred P. Sloan Foundation and Research Corporation for support. \end{acknowledgments}

\appendix
\section{King/NFW profiles}
\label{twob}
The King profile is parameterized by the expression
\begin{equation}
\rho(r)=\frac{9\sigma^{2}}{4\pi G a} \frac{1}{(1+\frac{r^{2}}{a^{2}})^{3/2}}\label{king},
\end{equation} where $\sigma$ is the cluster velocity dispersion, $a$ is its core radius, and $r$ denotes distance from the cluster center. The surface density for a King profile is derived by integrating by Eq.~(\ref{king}) along the line of sight, and is given by 
\begin{equation}
\Sigma(R)=\frac{9\sigma^{2}}{2\pi G a} \frac{1}{1+\frac{R^{2}}{a^{2}}}\label{kingsurface},\end{equation} where $R$ is the projected radius \cite{bt}. The projected mass density associated with the NFW mass profile,
\begin{eqnarray}
\nonumber
\rho(r)=\frac{\rho_{s}}{\left(\frac{r}{r_{s}}\right)\left(1+\frac{r}{r_{s}}\right)^{2}},\\
\rho_{s}=\frac{200c_{\rm NFW} ^{3}\rho_{{\rm crit}}}{3\left[\ln{(1+c_{\rm NFW})}-\frac{c_{\rm NFW}}{1+c_{\rm NFW}}\right]},\end{eqnarray}
is $\Sigma=r_{s}\rho_{s} f(x)$, where
\begin{eqnarray}
f(x)=
\left\{
\begin{array}{ll}
\frac{2\left\{1-\frac{2}{\sqrt{1-x^2}}\operatorname{arctanh}\left[\left(\frac{1-x}{1+x}\right)^{1/2}\right]\right\}}{x^{2}-1},&\mbox{if $x<1$};\\
\frac{2}{3},&\mbox{if $x=1$};\\
\frac{2\left\{1-\frac{2}{\sqrt{x^2-1}}\arctan{\left[\left(\frac{x-1}{x+1}\right)^{1/2}\right]}\right\}}{x^{2}-1},&\mbox{if $x>1$},\\
\end{array}\label{nfwsurface}
\right.
\end{eqnarray} $c_{\rm NFW}$ is the NFW concentration parameter, and $x=R/r_{s}$  \cite{nfw1,nfw2,nfw3,brainerd}.

\section{The effect of updated cluster mass-profiles on constraints obtained from A1413, A2256, and A2218.}
\renewcommand{\thetable}{B-\arabic{table}} 
\setcounter{table}{0}
\label{assume}
The values of $\sigma$ and $a$ used in Refs. \cite{ressellt,Bershady:1990sw} are shown in Table \ref{bershadytable}, along with the relevant redshift values and spectral slit locations. In Refs. \cite{ressellt,Bershady:1990sw}, the sky background was removed by subtracting `off' cluster spectra from `on' cluster spectra. In general, the expected signal due to axion decay, in the observer's frame, is
\begin{equation}
I_{\lambda_{0}}=\frac{\Sigma_{{\rm a}}(R)c^{3}}{4\pi \sqrt{2\pi} \sigma \lambda_{{\rm a}}\tau_{{\rm a}}(1+z_{{\rm cl}})^{4}}e^{-\frac{(\lambda_{0}/(1+z_{{\rm cl}})-\lambda_{{\rm a}})^{2}}{\lambda_{{\rm a}}^{2}}\frac{c^{2}}{2\sigma^{2}}}.\label{gen}
\end{equation} This can be shown by the same arguments used to derive Eq.~(\ref{lintens}). Using this ratio, we can figure out the ratio in expected signals. Since $I_{\lambda_{0}}\propto \xi^{2}$, we can obtain an estimate of the upper limit implied by the results of Refs. \cite{ressellt,Bershady:1990sw}, given current measurements of the cluster mass-profile and cosmological parameters. 

\begin{table}
\caption{Summary of observations and properties of clusters used in Refs. \cite{Bershady:1990sw,ressellt}. Table entries taken from Ref. \cite{Bershady:1990sw}.}
\begin{ruledtabular}
\begin{tabular}{ccccr}
&$\sigma$&$a$&Inner/Outer aperture&\\
 Cluster   &$({\rm km}~{\rm s}^{-1})$&$[{\rm kpc}({\rm arcmin})]$&$(R/a)$&$z$\\ \hline
 A1413&$1230$&$400h_{50}^{-1}(2.03)$&$1.11/4.64$&$0.143$\\
& & &$0.65/2.94$&\\ \hline
 A2218&$1300$&$200h_{50}^{-1}(0.88)$&$0.94/5.33$&$0.171$\\ \hline
 A2256&$1300$&$473h_{50}^{-1}(5.0)$&$0.484/2.96$&$0.0601$\\ \hline
\end{tabular}
\end{ruledtabular}
\label{bershadytable}
\end{table}

For A1413, we took best-fit values from the XMM-Newton x-ray profiles of Ref. \cite{a1413mass}, where it was found that A1413 is fit much better by an NFW profile than by a King profile. The best-fit NFW parameters are $c_{\rm NFW}=5.82$ and $r_{200}=r_{s}c_{\rm NFW}=1707~{\rm kpc}$. 
We use $\Omega_{{\rm m},{\rm new}}h_{\rm{new}}^{2}=0.15$, while $\Omega_{{\rm m},\rm{old}}h_{\rm{old}}^{2}=0.25$ is the value used in Refs. \cite{ressellt,Bershady:1990sw}. The projected mass density in axions is $\Sigma_{\rm{a}}=\left[\Omega_{\rm{a}}h^{2}/(\Omega_{{\rm m}}h^{2})\right]\Sigma$. We define an on-off density-contrast $\tilde{\Sigma}_{\rm{a}}^{\rm{new}}\equiv\Sigma_{\rm{a}}^{\rm{new}}(R_{\rm{on}})-\Sigma_{\rm{a}}^{\rm{new}}(R_{\rm{off}})$ using the best-fit values today. We define another,
 $\tilde{\Sigma}_{\rm{a}}^{\rm{old}}\equiv\Sigma_{\rm{a}}^{\rm{old}}(R_{\rm{on}})-\Sigma_{\rm{a}}^{\rm{old}}(R_{\rm{off}})$, using the best-fit values assumed in Refs. \cite{ressellt,Bershady:1990sw}. When calculating $\Sigma_{\rm{new}}$ at the slit locations of Refs. \cite{ressellt,Bershady:1990sw}, we took the slit locations in angular units and obtained physical distances using the angular-diameter distance for a $\Lambda$CDM cosmology. Applying Eq.~(\ref{nfwsurface}), we obtained $\tilde{\Sigma}_{\rm{a}}^{\rm{new},1}/\tilde{\Sigma}_{\rm{a}}^{\rm{old},1}=0.9853$ and $\tilde{\Sigma}_{\rm{a}}^{\rm{new},2}/\tilde{\Sigma}_{\rm{a}}^{\rm{old},2}=1.449$. Using Eq.~(\ref{gen}), it can be seen that this implies $\xi_{\rm{new},1}=1.104\xi_{\rm{old},1}$ and $\xi_{\rm{new},2}=0.831\xi_{\rm{old},2}$ for A1413.

The optical depth to lensing by A2256 is very low, because of the low redshift of the cluster. As a result, lensing derived mass models of this cluster do not exist. We took best-fit values from the BeppoSAX x-ray profiles of Ref. \cite{a2256mass}, in which King profiles are parameterized via \footnote{The factor $\rho_{s}$ usually appears in NFW profiles, and its use in a King profile is unusual, but correct.}\begin{eqnarray}
\Sigma^{\rm{new}}(R)=\frac{r_{c}\rho_{s}}{1+\frac{R^{2}}{r_{c}^{2}}}.\label{kingnew}
\end{eqnarray} We then used Eqs.~(\ref{kingnew}) and (\ref{kingsurface}) to obtain the ratio of the 
best-fit on-off density contrast determined using current data to that used in Refs. \cite{ressellt,Bershady:1990sw}:
 \begin{equation}
 \begin{array}{c}
 \frac{\tilde{\Sigma}_{\rm{a}}^{\rm{new}}}{\tilde{\Sigma}_{\rm{a}}^{\rm{old}}}=\frac{50 a r_{c} c_{\rm NFW}^{3} H^{2}}{9\sigma^{2}\left\{\ln{(1+c_{\rm NFW})}-\frac{c_{\rm NFW}}{1+c_{\rm NFW}}\right\}}\left[\frac{\Omega_{{\rm m},\rm{old}}h_{\rm{old}}^{2}}{\Omega_{{\rm m},\rm{new}}h_{\rm{new}}^{2}}\right]\\
 \times \frac{\left[\frac{1}{1+\left(\frac{a}{r_{c}}\right)^{2}\left(\frac{R_{\rm{on}}}{a}\right)^{2}}-\frac{1}{1+\left(\frac{a}{r_{c}}\right)^{2}\left(\frac{R_{\rm{off}}}{a}\right)^{2}}\right]}{\left[\frac{1}{1+\left(\frac{R_{\rm{on}}}{a}\right)^{2}}-\frac{1}{1+\left(\frac{R_{\rm{off}}}{a}\right)^{2}}\right]}\end{array}.
 \label{em2256}
  \end{equation} Here, $H$ is the value of the Hubble constant preferred today. Using BeppoSax data, best-fit values of  $c_{\rm NFW}=4.57$ and $r_{c}=570~\rm{kpc}$ were derived in Ref. \cite{a2256mass}, using a redshift of $z=0.0581$, and assuming a sCDM cosmology. Rescaling this core radius for a $\Lambda$CDM universe, we obtain $r_{c}=414~\rm{kpc}$. Inserting these values into Eq.~(\ref{em2256}), we obtain
$\tilde{\Sigma}_{\rm{a}}^{\rm{new}}/\tilde{\Sigma}_{\rm{a}}^{\rm{old}}=0.5982$. For A2256, this yields $\xi_{\rm{new}}=1.29\xi_{\rm{old}}$. If true, recent claims that A2256 is undergoing merging activity impugn the assumption that A2256 is relaxed \cite{1991A&A...246L..10B,a2256dyn2}. In that case, the assumption of a King profile for A2256 is invalid, and upper limits to $\xi$ obtained from A2256 have to be revised.

The strong-lensing analyses of A2218 in Refs. \cite{a2218mass,smithpape} indicate the presence of several mass clumps in the cluster, four of which have total masses comparable to the total cluster mass, and one, centered on the brightest cluster galaxy (BCG), which has a total mass comparable to a typical galaxy mass. The observed lensing configuration is well fit by the set of parameters listed in Table \ref{2218m}. The parameters refer to a PIEMD \cite{kassiolakovner1993}, whose surface mass density is given by
\begin{eqnarray}
\nonumber
\Sigma(x,y)=\frac{\sigma_{0}^{2}}{2G}\frac{r_{\rm{cut}}}{r_{\rm{cut}}-r_{\rm{core}}}\left[
\frac{1}{\left(r_{\rm{core}}^{2}+s^{2}\right)^{1/2}}\right].\\
\nonumber
s^{2}=\left[\frac{x-x_{c}}{1+\epsilon}\right]^{2}+\left[\frac{y-y_{c}}{1-\epsilon}\right]^{2},\\
\epsilon=\frac{a/b-1}{a/b+1},
\end{eqnarray} where $a$ and $b$ are the semi-major and semi-minor axes of the best-fitting ellipse, $x_{c}$ and $y_{c}$ are the best-fitting mass centers given in Table \ref{2218m}, translated into physical units using the $\Lambda$CDM angular-diameter distance, and $\sigma_{0}$ is the velocity dispersion of the cluster. Although these lensing data were analyzed using a sCDM cosmology, the authors report that the best-fit parameters are insensitive at the $10\%$ level to reasonable variations in cosmological parameters.
\begin{table}
\caption{Best-fit parameters for the mass model of A2218, determined from a strong-lensing analysis. The table was taken from Refs. \cite{a2218mass,smithpape}. Square brackets indicate a value that was not fit for, but set by hand\label{masst}. The quantity $\theta$ is the orientation of the ellipse's major axis relative to some horizontal in the image plane.}
\begin{ruledtabular}
\begin{tabular}{rrrrrrrr}
$\Delta\mbox{R.A.(")}$ &$\Delta\mbox{Dec.(")}$&$a/b$&$\theta$(deg)&$r_{\rm{core}}$(kpc)&$r_{\rm{cut}}$(kpc)&$\sigma_{0}$\\ \hline
$+0.2$&$+0.5$&$1.2$&$32$&$83$&$\left[1000\right]$&$1070$\\ \hline
$\left[+47.0\right]$&$\left[-49.4\right]$&$1.4$&$53$&$57$&$\left[500\right]$&$580$\\ \hline
$\left[+16.1\right]$&$\left[-10.4\right]$&$\left[1.1\right]$&$\left[70\right]$&$<2$&$65$&$195$\\ \hline
$\left[4.8\right]$&$\left[-20.9\right]$&$\left[1.4\right]$&$\left[-23\right]$&$<2$&$77$&$145$\\ \hline
$+0.3$&$+0.1$&$1.8$&$53$&$<3$&$136$&$270$\\ \hline
\end{tabular}
\end{ruledtabular}
\label{2218m}
\end{table}
The on-off radii are provided without orientation information in Refs. \cite{ressellt,Bershady:1990sw}, and so we allow the slit orientation angle $\phi$ to vary over the full possible range, and repeat the preceding analysis to obtain a range $0.57\xi_{\rm{old}}\leq \xi_{\rm{new}}\leq0.71\xi_{\rm{old}}$. Parameters whose values are bounded from above are set to zero for our analysis. Depending on the mass bin, the upper limits of Refs. \cite{ressellt,Bershady:1990sw} come from A2256 or A2218. The updated upper limits of Refs. \cite{ressellt,Bershady:1990sw} must thus fall in the bracket $0.57\xi_{\rm{old}}\leq\xi_{\rm{new}}\leq 1.29\xi_{\rm{old}}$.
This range is plotted in Fig.~\ref{fluxxilimitbin}, and it is clear that our upper limits to $\xi$ improve considerably on those reported in Refs. \cite{ressellt,Bershady:1990sw}, even when past work is reinterpreted optimistically.

\end{document}